\documentclass{llncs}
\usepackage{graphicx}
\usepackage[utf8]{inputenc}
\usepackage{multicol}
\usepackage{multirow}
\usepackage{url}
\begin{document}
\title{Constraint-Guided Workflow Composition Based on the EDAM Ontology}
%
\titlerunning{Constraint-Guided Workflow Composition}
\author{Anna-Lena Lamprecht\inst{1} \and Stefan Naujokat\inst{1} \and Bernhard Steffen\inst{1} \and Tiziana Margaria \inst{2}}
\authorrunning{Lamprecht et al.}   
\institute{Technical University Dortmund, Chair for Programming Systems, Dortmund, D-44227, Germany\\
\email{\{anna-lena.lamprecht|stefan.naujokat|bernhard.steffen\}@cs.tu-dortmund.de}
\and
University Potsdam, Chair for Service and Software Engineering, Potsdam, D-14482, Germany\\
\email{tiziana.margaria@cs.uni-potsdam.de}}

\maketitle              
\begin{abstract}
Methods for the automatic composition of services into executable workflows
need detailed knowledge about the application domain, in particular
about the available services and their behavior in terms of input/output data descriptions.
In this paper we discuss how the EMBRACE data and methods ontology (EDAM) can be used as
background knowledge for the composition of bioinformatics workflows.
We show by means of a small example domain that the EDAM knowledge facilitates
finding \emph{possible} workflows, but that additional knowledge is required to guide the
search towards actually \emph{adequate} solutions. 
We illustrate how the ability to flexibly formulate domain-specific and problem-specific constraints
supports the workflow development process.
\end{abstract}

\section{Introduction}
The challenge of automatic workflow composition, in particular the automatic composition of bioinformatics
web services into executable workflows, has been addressed by several projects in the past years 
(see, e.g., \cite{DiPoWi2008,LaMaSt2009,WiVaMc2009,RiKaTr2009,MaRGRT2010}).
The importance of domain-specific knowledge about services and data types has been recognized
long ago (see, e.g.,\cite{CSGTCP2003}). However, most automatic workflow composition systems have so far relied
on self-defined, special-purpose domain models, rather than using systematic information from a central instance
(simply because no such instance existed).
Some of the systems named above do in fact use knowledge from the BioMoby \cite{WilLin2002} ontologies (which apply the
LSID \cite{ClMaLi2004} naming scheme), but these have mainly been derived from the (often incomplete or imprecise) meta-information
that service providers submit during service registration at MobyCentral, and not systematically designed for depicting the structure
of the whole bioinformatics domain. 
Quite recently the EDAM (EMBRACE Data And Methods) ontology \cite{PIKTMJ2010} has been 
initiated with the aim of building a unified vocabulary about bioinformatics services and data, 
suitable for bridging the gap between mere service registries and semantically aware service composition methodologies.


In this paper we show how the domain knowledge that EDAM provides can be used as basis for the
automation of service composition. 
After an introduction to the principles of automatic workflow composition and to the
software framework that we used for this study (Section \ref{sec:poawc})
we describe the EDAM ontology (Section \ref{sec:edam}), especially focussing on the parts that are 
relevant for automatic workflow composition.
In Section \ref{sec:examples} (Results and Discussion) we present an example domain and workflow composition problem,
showing that the EDAM knowledge facilitates finding \emph{possible} workflows, but that additional knowledge
is required to limit the search to the actually \emph{desired} or \emph{adequate} solutions.
We use the PROPHETS synthesis framework, which facilitates a very flexible way of expressing additional knowledge:
it can either be specified during domain modeling (especially suitable for domain-specific
constraints) or during the actual synthesis (especially suitable for problem-specific constraints).
In this way, users can flexibly interact with the workflow development framework, collecting possible solutions
and continuously refining the constraints according to any arising requirements.
The paper ends with a conclusion in Section \ref{sec:conclusion}.


\section{Automatic Workflow Composition}
 \label{sec:poawc}

The automatic composition of (small) software units into (large) runnable pieces of software has been 
subject to research for several years (see, e.g. 
\cite{McCHay1969,FikNil1971,ManWol1984,PnuRos1989,ErHeNa1994,StMaBe1997,KupVar2000,MaOrBL2000,MarSte2007}).
Contemporary terminology, for instance \emph{automatic workflow composition} or \emph{automatic service composition},
reflects the current trend towards service orientation and (business) process modeling.
Nevertheless, the principles of the composition methodologies on which they are based remain the same.

The commonly known methods can roughly be distinguished into synthesis-based and planning-based approaches:
Synthesis methods usually are rather behavior-oriented (e.g., the
temporal-logic-based methods of \cite{PnuRos1989,KupVar2000}), whereas
classical planning algorithms are more focused on the availability of resources and
definition of a world state with predicates, which can be modified by the
actions (cf. \cite[chapter 10]{RusNor2009}). 
In addition, several hybrid approaches, such as LTL-enhanced planning
\cite{BacKab2000,MaOrBL2000}, exist that incorporate aspects of the
respective other method.
Despite all the differences with regard to the above named strategies and algorithms,
all approaches share an essential characteristic: they search for paths in some kind of ``universe" that
is given by the information about the services and in particular their input/output behavior.

Accordingly, the quality of the solutions crucially depends on the quality of the available information
about the services of the domain (more than on the actual search strategy): 
Merely syntactic information (as provided by most programming language APIs or standard web service
interfaces) is not sufficient. Proper semantic descriptions of services and data types are required
to obtain meaningful results.
For instance, assume a web service that returns a job ID, and another
service that consumes a nucleotide sequence. From a programming language point of view, both would
be classified as character sequences (strings) and would thus be assumed to match. While this is 
syntactically correct, submitting a job ID to a sequence analysis service would of course not lead
to the desired results, so a more precise - semantic - interface description is required.

\subsection*{Automatic Workflow Composition in Bio-jETI}

For the work presented in this paper, we used the PROPHETS extension of the Bio-jETI framework \cite{MaKuSt2008}.
PROPHETS seamlessly integrates automatic service composition methodology
into Bio-jETI, in particular by supporting visualized/graphical semantic domain modeling,
loose specification within the workflow model, non-formal specification of constraints using 
natural language templates, and automatic generation of model checking formulas (to check global
properties of workflows).
For a detailed introduction into the framework and the underlying ideas the reader is referred to 
\cite{LaNaMS2010,LaNaMS2010a}. In a nutshell, working with PROPHETS incorporates two phases:
\begin{enumerate}
\item During \emph{domain modeling} meta-information about the available services is collected and 
provided in appropriate form, and semantic classifications of the services and their input and output types
are defined. The service and type
taxonomies are stored in OWL format, where the OWL classes represent abstract classifications,
and the actual types and services are then represented as individuals that are related to one or more of 
those classifications by \emph{instance-of} relations. 
\item In the actual \emph{workflow design} phase this domain is used to (automatically) create the 
workflow models. Central is the idea of \emph{loose specification}: branches in the model can be 
marked as \emph{loosely specified}, which are then automatically replaced by adequate composite services. 
\end{enumerate}

The algorithm \cite{StMaFr1993} that is then applied to complete a loosely specified
workflow to be fully executable takes two orthogonal aspects into account: On the
one hand, the workflow must be type-consistent to be executable, on
the other hand, the constraints specified by the workflow designer must be met:
\begin{itemize}
\item The \emph{configuration universe} constitutes the algorithm's basic search
space. It contains all valid execution sequences and is implicitly defined
by the domain model. 
As this
configuration universe is usually very large, it is not explicitly generated
from the domain definition, but on the fly during the synthesis process.
\item The \emph{specification formula} is the second aspect.
It describes all sequences of services that meet the individual workflow
specification, but without taking care of actual executability concerns. As the
explicit representation of all these sequences would be extremely
large, the formula is not explicitly built, but given as a declarative formula in
SLTL (Semantic Linear Time Logic) \cite{StMaFr1993}, a variant of the commonly known propositional
linear-time logic (PLTL). The formula is a conjunction of all available
constraints, comprising:
\begin{itemize}
\item technical constraints (configurations of the synthesis algorithm),
\item domain-specific constraints (globally specified by the domain modeler), and 
\item problem-specific constraints (derived from the loosely specified branch or specified by the workflow
designer).
\end{itemize}
\end{itemize}

To start the search for solutions, the synthesis algorithm requires an initial
state (i.e. a set of start types). They
are determined automatically according to the output specification of the service at the beginning of the 
loose specification. Based on this the synthesis algorithm performs a parallel evaluation
of the configuration universe and the specification formula to search for paths
that are consistent with both the configuration universe and the SLTL
formula.  Each of those paths is a valid concretization of the loosely specified branch.

\section{EDAM as Background Knowledge}
\label{sec:edam}

The EDAM (EMBRACE Data And Methods) ontology\footnote{\url{http://edamontology.sourceforge.net/}} 
has been developed in the scope of the EMBRACE (European Model for Bioinformatics Research and
Community Education) project\footnote{\url{http://www.embracegrid.info/}}
as an ontology for describing life science web services \cite{PIKTMJ2010}.
In contrast to many known ontologies like the Gene Ontology \cite{ABBBBC2000} or the majority of the Open 
Biomedical Ontologies \cite{SARBBC2007},
who focus on the description of biological content,
it provides a vocabulary of terms and relations that can be used for annotating
services with useful (semantic) metadata, in particular regarding their behavior and their inputs and outputs. 
Two important applications of EDAM have already been identified in \cite{PIKTMJ2010}: 
the use of the defined terms for semantic annotations of web services (e.g. via SAWSDL
extension attributes \cite{FarLau2007}) in order to facilitate service discovery and integration,
and the more detailed description of the involved data types in order to improve
the data exchange between services.

Strictly speaking, EDAM is not a single, large ontology, but consists of six separate (sub-) ontologies:
\begin{enumerate}
\item \textbf{Biological entity:} physically existing (parts of) things, such as \emph{SNPs}, \emph{alleles}, or \emph{protein domains}. 
\item \textbf{Topic:} fields of bioinformatics study, such as \emph{nucleic acid sequence analysis}, \emph{model organisms}, or \emph{visualization and rendering}.
\item \textbf{Operation:} particular functions of tools or services (e.g., web service operations), such as \emph{annotation}, \emph{codon usage analysis},
or \emph{sequence database search by motif or pattern}. 
\item \textbf{Data resource:} various kinds of data resources, such as \emph{Biological pathways data resource}, \emph{Literature data resource}, 
or \emph{Ontology data resource}.
\item \textbf{Data:} semantic descriptions of data entities that are commonly used in bioinformatics, such as
\emph{BioCyc enzyme ID}, \emph{phylogenetic consensus tree}, or \emph{sequence alignment}.
\item \textbf{Data format:} references to (syntactic) data format specifications, such as \emph{Clustal sequence format},
\emph{PubMed Central article format}, or \emph{HMMER hidden Markov model format}.
\end{enumerate}

Importantly, EDAM is not a catalogue of concrete services, data, resources etc., but a provider of terms for 
classification of such entities -- exactly what is meant by ``background knowledge".
Thus, using EDAM is simply taking the (relevant parts of the) ontology and sorting the application-specific resources
into this skeletal domain structure.
In the following we use the \emph{Operation} (sub-) ontology as vocabulary for service classifications, 
and the \emph{Data} 
(sub-) ontology for data type classifications.
Employing also the \emph{Data format} part in order to take into account the more technical interface
specifications is subject of ongoing work.
(For this study we used EDAM version beta08.)



\section{Results and Discussion}
\label{sec:examples}

Using EDAM as background knowledge in the domain model for synthesis with PROPHETS, 
the domain setup involves three major steps:
\begin{enumerate}
\item Converting EDAM from OBO (Open Biomedical Ontologies) format into OWL format (using the Protégé OWL API).
\item Generating the service taxonomy from the \emph{Operation} term and (transitively) all its subclasses, and 
the type taxonomy from the \emph{Data} term and (transitively) all its subclasses.
\item Sorting the available services and their input/output types into the service and type taxonomy, respectively.
\end{enumerate}
Steps 1 and 2 can be executed fully automatically.
Step 3 can be automated if EDAM annotations are available for the types and services, 
as it is the case, for instance, for EMBOSS \cite{RiLoBl2000}, which provides EDAM relations in its Ajax Command Definition (ACD) files.
In the following we will use a smaller example domain that comprises a set of bioinformatics tools for which the taxonomic classifications
have been defined manually. It uses only a small part of EDAM, but it is already large enough to illustrate our results. 

\begin{table}[htdp]
\caption{Domain model: service descriptions.}
\begin{center}
\begin{tabular}{|l|l|}
\hline
\textbf{Service}							&	\textbf{Behavior} \\
\hline
\hline
ClustalW								&	In: \textit{Sequence}			\\
\textit{Global multiple sequence alignment}	& 	Out: \textit{Multiple sequence alignment}	\\
\hline
ClustalW2								&	In: \textit{Sequence}		\\
\textit{Global multiple sequence alignment}	&	Out: \textit{Multiple sequence alignment}	\\
\hline
DBFetch\_FetchBatch					&	In: \textit{Sequence identifier}	\\
\textit{Database search and retrieval}		&	Out: \textit{Sequence}	\\
\hline
DBFetch\_FetchData					&	In: \textit{Sequence identifier}			\\
\textit{Database search and retrieval}		&	Out: \textit{Sequence}	\\
\hline
Gblocks								&	In: \textit{Multiple sequence alignment}	\\
\textit{Sequence alignment conservation analysis}	& Out: \textit{Multiple sequence alignment}				\\
\hline
KAlign								&	In: \textit{Sequence}		\\
\textit{Global multiple sequence alignment}	&	Out: 	\textit{Multiple sequence alignment}				\\
\hline
Mafft									&	In: \textit{Sequence}		\\
\textit{Global multiple sequence alignment}	&	Out: \textit{Multiple sequence alignment}		\\
\hline
Muscle								&	In: \textit{Sequence}		\\
\textit{Global multiple sequence alignment}	&	Out: \textit{Multiple sequence alignment}				\\
\hline
PhyML\_AminoAcid						&	In: \textit{Protein Sequence}	\\
\textit{Phylogenetic tree construction}		&	Out: \textit{Phylogenetic tree}	\\
\textit{from molecular sequences}			&					\\
\hline
PhyML\_DNA							&	In: \textit{DNA sequence}		\\
\textit{Phylogenetic tree construction}		&	Out:  	\textit{Phylogenetic tree}		\\
\textit{from molecular sequences}			&					\\
\hline
poptree\_NJ							&	In: \textit{Sequence composition}	\\
\textit{Phylogenetic tree construction}		&	Out:  poptree\_outfile	\\
\textit{(minimum distance methods)}			&							\\	
\hline
poptree\_UPGMA						&	In: \textit{Sequence composition}		\\
\textit{Phylogenetic tree construction}		&	Out: poptree\_outfile 	\\
\textit{(minimum distance methods)}			&				\\
\hline
postree								&	In: poptree\_outfile			\\
\textit{Phylogenetic tree drawing}			&	Out: \textit{Phylogenetic tree image}		\\
\hline
predator								&	In: \textit{Protein sequence}			\\
\textit{Protein secondary structure prediction}	&	Out: \textit{Protein secondary structure}				\\
\hline
ps2pdf								&	In: \textit{Image}			\\
									&	Out: \textit{Image}				\\
\hline
ReadFile								&	Out: 	\textit{Data}				\\
\textit{File loading}						&							\\
\hline
ReadDNASequence						& 	Out: \textit{DNA sequence}		\\
\textit{File loading}						&							 				\\
\hline
TCoffee								&	In: \textit{Sequence}			\\
\textit{Global multiple sequence alignment}	&	Out: \textit{Multiple sequence alignment}				\\
\hline
WriteFile								&	In: \textit{Data}					\\
\hline
WUBlast								&	In: \textit{Sequence}		\\
\textit{Sequence database search by sequence}&	Out: \textit{Sequence database hits}				\\
\textit{(word-based methods)}				&						\\
\hline
Viewer								&	In: \textit{Data}		\\
\textit{Visualisation and rendering}			&			\\
\hline
WUBlastParser							&	In: \textit{Sequence database hits}		\\
									&	Out: \textit{Sequence identifier}		\\
\hline
\end{tabular}
\end{center}
\label{table:domainmodel}
\end{table}%

Table \ref{table:domainmodel} lists the services in the example domain, along with their input and output
data types. Concrete service and data type names are given in normal font, while (abstract) services and types
in terms of the EDAM ontology are given in italics. 
Note that the service interface descriptions in this domain are quite simple: each one has at most one input and one output,
whereby only user data (i.e. input and output files) are considered, while other parameters (primarily used for configurations
that are not actually inputs) are not taken into account.

\begin{figure}
\begin{center}
\includegraphics[width=\textwidth]{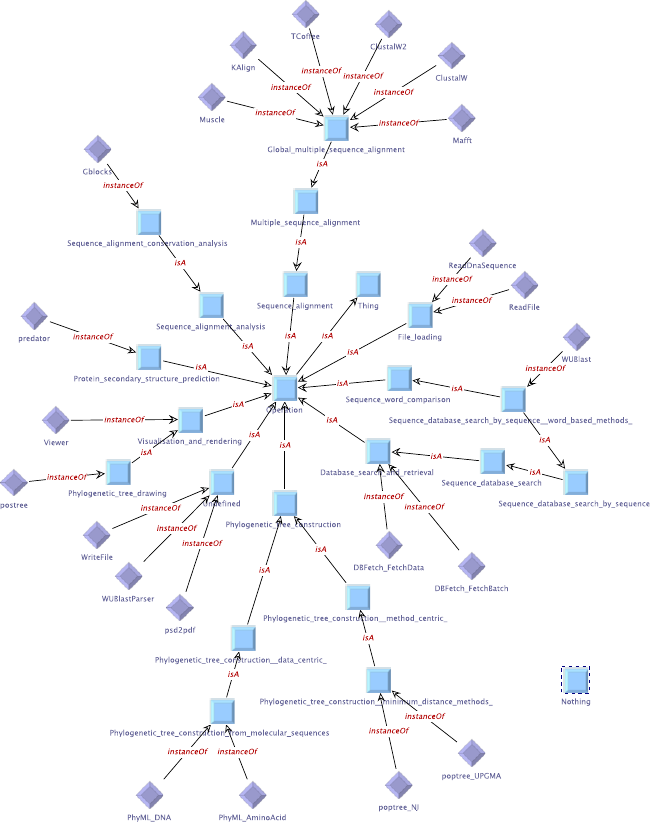}
\caption{Service taxonomy.}
\label{fig:servicetaxo}
\end{center}
\end{figure}

\begin{figure}
\begin{center}
\includegraphics[width=\textwidth]{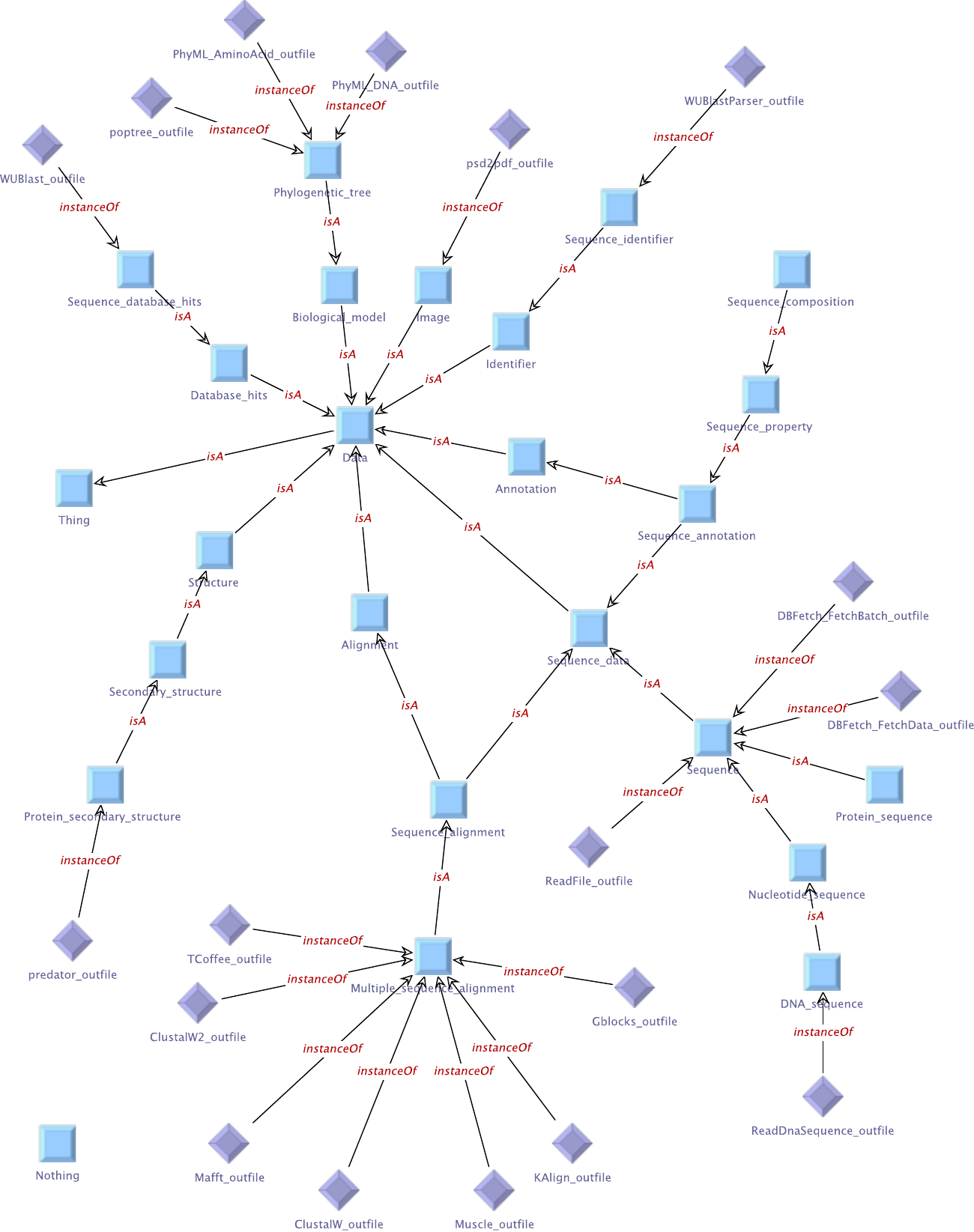}
\caption{Type taxonomy.}
\label{fig:typetaxo}
\end{center}
\end{figure}

Figures \ref{fig:servicetaxo} and \ref{fig:typetaxo} show the service and type taxonomies (screenshots from
PROPHETS' built-in ontology editor). 
The classes (the blue squares) correspond to the \emph{Operations} and \emph{Data} (sub-) ontologies of EDAM, respectively, but have
been cut down to the classes that are relevant for the services and data that are used in the domain model.
The purple rhombs represent the concrete services and types of the domain that have been added to the domain model
as instances of the respective EDAM classes.

\begin{figure}[htb]
\begin{center}
\includegraphics[width=\textwidth]{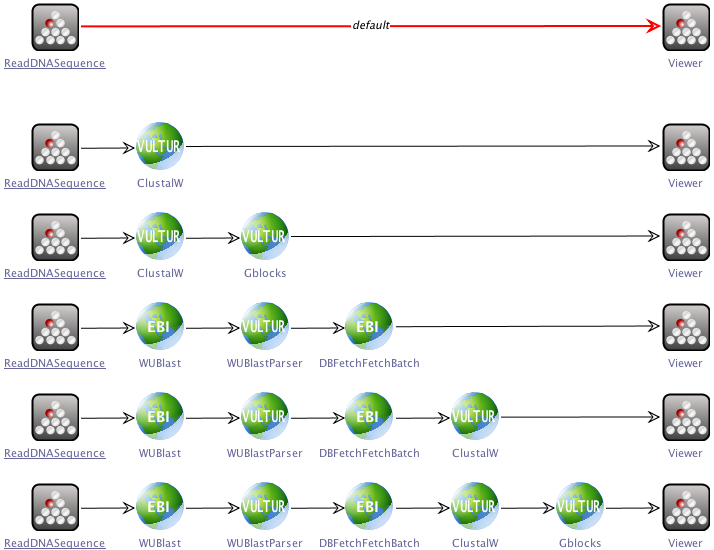}
\caption{Loosely specified workflow and some possible concretizations.}
\label{fig:workflow}
\end{center}
\end{figure}

Figure \ref{fig:workflow} shows the simple loosely specified workflow that we used as basis for experimentation with the 
described domain model, along with five possible concretizations. The workflow begins with reading a DNA sequence file 
(ReadDNASequence) and ends with displaying a result (Viewer). These components are connected by a loosely
specified branch (colored in red). Accordingly, the basic synthesis problem is to find a workflow that takes a ReadDNASequence's output
(a DNA sequence) as input, and produces Viewer's input (some data) as output.
Note that this setup is in particular appropriate for experimentation with the domain, exploring what workflows are generally
possible with certain input data. In many realistic applications, the target of the loosely specified branch is a more specific component,
for instance an alignment viewer, so that the synthesis problem as such becomes more specific, too.

The five concretizations shown in Figure \ref{fig:workflow} are of course not the only possible ones.
As we will detail in the following, thousands if not millions of 
solutions are easily \emph{possible} with the described domain model, but they are not necessarily \emph{desired}  or \emph{adequate}.
We will show in the following how playing with configurations and constraints may help mastering this enormous potential by excluding 
inadequate or bundling equivalent solutions.

In a first experiment, we executed the synthesis with no further constraints. That is, only the input/output specification
defined by the loosely specified branch was considered and a ``naive" search was performed on the synthesis universe.
When we used the basic configuration of the synthesis algorithm, we obtained 264,118 possible solutions for
the synthesis problem already for a search depth of 5. 
Additionally using a solution filtering mechanism that removes service sequences that are mere permutations of others decreased
the number of solutions to 5,325.
As the example domain is characterized by consisting of services that have at most one input and one output data type,
it turned out to be most efficient, however, to use a configuration in which the synthesis also mimics a ``pipelining'' behavior. This means that output
data is only transferred to the direct successor of a service in the workflow and in particular not available for any other subsequent services.
This configuration led to 2269 solutions in the unconstrained case. 


In the following we will develop a small set of constraints that is already sufficient to drastically 
reduce the solution space further by excluding clearly inadequate solutions.
The constraints are based on a few general observations about the solutions obtained so far:

\begin{enumerate}
\item Workflows that contain services that make \textbf{no contribution} to solving the synthesis problem. Such services are,
for example, ReadFile (requiring no input but producing arbitrary new data that is planted into the workflow and can distract
from the actual synthesis problem) or WriteFile (consuming data but without producing a new analysis result).
\item Workflows that contain services that make \textbf{no progress} within the workflow. Particular examples are redundant service calls,
such as a the multiple invocation of Gblocks on a multiple sequence alignment where only the first call does make sense.
\item Workflows that contain \textbf{``dead" functionality} in the sense that certain service outputs are not adequately used. For instance,
a BLAST result is usually not used as such. Rather, the sequence identifiers of the BLAST hits are of interest in order to retrieve the
corresponding entries from a database. Thus, to make sense, workflows that call BLAST should also contain a call to a BLAST parser that gets
the IDs from the BLAST result.
\item Workflows that contain several services that are useful when seen individually or in parts of the solution, but the overall workflow
is \textbf{not the envisaged analysis}.  For example, a solution where an alignment service (such as ClustalW) is called with the input data
is certainly a possible and useful workflow, but if the workflow developers' intention was actually to do the alignment with homologous
sequences of the input sequences, or to obtain a phylogenetic tree from the input, this solution is not adequate.
\end{enumerate}

In order to address the first observation (``services that make no contribution'') we defined a constraint that excludes
all corresponding services  of our example domain (i.e. ReadFile, ReadDNASequence, WriteFile, and Viewer) from the synthesis.
This can easily be realized via the SLTL formula $$G(\neg \langle \texttt{S} \rangle true)$$ which prohibits any occurrence
of $\texttt{S}$ in a solution. Since the exclusion of particular services is a quite common constraint, it is provided 
as a template in the PROPHETS Synthesis Wizard, which only requires the user to list all `unwanted' services.
Applying this template to all four services named above yields a constraint that in itself reduced the number of 
solutions from 2,269 to 55.

Adding an SLTL constraint for the second observation (``services that make no progress'') in a similar fashion, 
specifying that Gblocks should not never be called twice (or more), decreases the number of solutions 
49. This number can be further reduced by an SLTL constraint exploiting the third observation (``dead functionality''):
requiring that a call of WUBlast should be followed by a call to the corresponding parser eliminates 
another 18 solutions.

In order to address the last observation (``workflow is not the envisaged analysis''), we provided the synthesis with constraints that
express general ideas about the desired workflow.
As one example, a frequently performed analysis for molecular sequences is to search for similar sequences in a database and
compare the new sequences to the known ones by means of a sequence alignment.
Accordingly, we formulated a constraint that enforces
a \emph{sequence database search by sequence} and after that a \emph{multiple sequence alignment} as part of the solution.
This decreased the number of solutions further down to 24.
The remaining solutions are similar to the two workflows at the bottom of Figure \ref{fig:workflow}:
they start with WUBlast and the WUBlastParser, followed by either DBFetch\_FetchBatch or
DBFetch\_FetchData, after which one of the alignment algorithms is called. Half of the solutions appended Gblocks as
final service.

Another analysis that is often applied to molecular sequences is the construction of a phylogenetic tree, which
represents the evolutionary relationship between the sequences. 
As an alternative to the above constraint for the fourth observation, we formulated a constraint that enforces the 
use of a \emph{phylogenetic tree construction} service. 
The only remaining result when this constraint is applied together with constraints 1, 2 and 3  is a one-step solution consisting of 
the PhyML\_DNA service, which computes a phylogenetic tree
from a set of DNA sequences.

\begin{table}[htdp]
\caption{Overview of results (considering solutions found until a search depth of 5).}
\begin{center}
\begin{multicols}{2}
\begin{tabular}{|c|r|r|}
\hline
\textbf{Constraints}	&	\textbf{Visited nodes}	&	\textbf{Solutions} \\
\hline
none					&		34,026~			&		2,269~		\\
1					&		1,139~			&		55~			\\
2					&		82,343~			&		2,194~		\\
3					&		132,809~			&		1,916~		\\
4					&		436,102~			&		471~			\\
4'					&		129,200~			&		406~			\\
1, 2					&		1,103~			&		49~			\\
1, 3					&		3,123~			&		52~			\\
1, 4					&		8,309~			&		24~			\\
1, 4'					&		2,336~			&		1~			\\
2, 3					&		138,137~			&		1,847~		\\
2, 4					&		443,860~			&		459~			\\
2, 4'					&		181,365~			&		394~			\\
3, 4					&		910,672~			&		138~			\\
3, 4'					&		277,239~			&		359~			\\
4, 4'					&		847,845~			&		18~			\\
\hline
\end{tabular}

\begin{tabular}{|c|r|r|}
\hline
\textbf{Constraints}	&	\textbf{Visited nodes}	&	\textbf{Solutions} \\
\hline
1, 2, 3				&		9,603~			&		31~			\\
1, 2, 4				&		8,057~			&		24~			\\
1, 2, 4'				&		2,084~			&		1~			\\
1, 3, 4				&		28,545~			&		24~			\\
1, 3, 4'				&		18,699~			&		0~			\\
1, 4, 4'				&		15,919~			&		0~			\\
2, 3, 4				&		919,162~			&		138~			\\
2, 3, 4'				&		284,463~			&		347~			\\
2, 4, 4'				&		859,047~			&		18~			\\
3, 4, 4'				&		1,752,153~		&		0~			\\
1, 2, 3, 4				&		28,545~			&		24~			\\
1, 2, 3, 4'				&		2,084~			&		1~			\\
1, 2, 4, 4'				&		15,235~			&		0~			\\
1, 3, 4, 4'				&		54,711~			&		0~			\\
2, 3, 4, 4'				&		1,764,843~		&		0~			\\
all					&		54,027~			&		0~			\\
\hline
\end{tabular}
\end{multicols}
\end{center}
\label{table:results}
\end{table}%

Table \ref{table:results} summarizes our findings, containing also all other combinations of the constraints developed above:
The first constraint (excluding services that make no contribution towards solving the synthesis problem) has the
strongest impact on the number of solutions, whereas the second constraint leads only to an exclusion of around 80 solutions.
Notably, already constraints 1 and 4 together decrease the number of solutions to 24, i.e. to the
set of solutions that was also obtained with the first four constraints together.
Likewise, constraints 1 and 4' together already lead to the single-step solution that was initially obtained by a combination
of four constraints.

As the table furthermore shows, some combinations of constraints, in fact most combinations involving 4 and 4',  
leave no solutions at all. This may indicate that the constraints are inconsistent (i.e. do not have any 
solution), or, as in this case, that they require solutions of length greater than the current search depth of 5. 
In the latter case, increasing the search depth may be the solution of choice, but there are two options which
apply in both cases:
\begin{itemize}
\item loosening the constraints, in order to resolve the inconsistencies or to allow shorter solutions, and 
\item revise/extend the domain model in order to increase the solution space.
\end{itemize}
Whereas playing with the first option may well be in the competence of the workflow designer, the second option 
requires domain modeling expertise.


The second column of the table gives the number of nodes that are visited by the synthesis algorithm during its
iterative deepening depth-first search. The numbers reflect that the synthesis search space is not only constituted by the
(static) service descriptions, but also by the additionally provided logical constraints (cf. Section \ref{sec:poawc}).
In fact, the search space may grow with the product of the sizes of the formula and the constraints. 
As the table shows, in our case constraints can both decrease and increase the search space: In the unconstrained case, 34,026 nodes
are visited, the least number of nodes (1,103) is visited when constraints 1 and 2 are used, and with constraints
2, 3, 4, and 4' a total of 1,764,843 nodes are visited until a search depth of 5.

Note that not all constraints we defined here are useful in every setting.
For instance, services that generate data ``from scratch"
may well be useful when there are other services involved that need this additional 
input. Thus, in cases where producing entirely new data is considered a useful or even 
necessary feature, the first constraint needs to be relaxed. Thus changing the considered
problem/purpose may in addition to adapting the purpose-specific constraints (here 4 and 4') 
also require to reconsider all the other constraint classes.


\section{Conclusion}
\label{sec:conclusion}

Methods for the automatic composition of services into executable workflows
need detailed knowledge about the application domain.
In this paper we discussed how the EMBRACE data and methods ontology (EDAM) can be used as
background knowledge for the composition of bioinformatics workflows.
We found that the EDAM knowledge facilitates
finding possible workflows, but that additional knowledge is required to limit the
search to the actually desired/adequate solutions:
\begin{itemize}
\item EDAM provides a controlled vocabulary for data and methods in the bioinformatics domain,
whereby it covers in particular the service and data type description terminology that is
needed for the automatic composition of services into workflows.
Given that the services in the domain and their input/output types are properly annotated in terms
of EDAM, this knowledge ensures that the synthesis algorithms find adequate workflows.
\item However, finding the actually desired workflows requires more knowledge.
In this paper we provided simple examples of additional constraints, such as exclusion of particular
services, dependencies between services, and general patterns for a desired solution.
\end{itemize}

Larger service collections like EMBOSS \cite{RiLoBl2000} or the BioCatalogue \cite{BTNLOR2010},
which provide hundreds or even thousands of services, are
not manageable without systematic discovery or service composition techniques.
When dealing with small domains, human experts may be unbeatable in composing tailored workflows,
but firstly not every human is an expert in bioinformatics services and data types,
and secondly also experts cannot always keep track of all changes in large domain libraries.
Thus both experts and average users may profit from tools that automatically exploit 
arising domain-specific and problem-specific knowledge beyond the usual 
``static" domain model (i.e. service interface descriptions).


The PROPHETS synthesis framework enables a very flexible way of expressing additional knowledge:
it can either be specified during domain modeling (especially suitable for domain-specific
constraints) or during the actual synthesis (especially suitable for problem-specific constraints).
Thus, the expression of (additional) domain knowledge is, in particular, cleanly separated from the implementation of the synthesis algorithm.
This is in contrast to other approaches to automatic composition of bioinformatics services that
we are aware of (such as \cite{DiPoWi2008,RiKaTr2009,MaRGRT2010}), which rely on the knowledge that is provided by the
service and data type descriptions and ontological classifications, and where all additional domain knowledge (if any)
is hidden in specifically designed composition algorithms.

Currently we are exploring the bioinformatics domain further in order to identify general domain-specific
constraints, and problem-specific constraints especially in the shape of often recurring workflow patterns.
Such a library of constraints that can be added and removed dynamically during the workflow development
process will enable users to work and experiment with the domain in a very flexible manner by tailoring
their solution space on demand. This kind of experimentation provides users with an easy entry into this 
complex landscape of tools and technologies, and later with means for scalability: experimenting with 
options and constraints it is possible to tailor the setting in a way that at the same time improves the 
adequacy of the specification as well as the search depths for solutions of the synthesis procedure.
We plan to support this approach by extending the domain knowledge with domain-specific search heuristics
enhancing the synthesis algorithms.


\bibliographystyle{splncs}
\bibliography{world}

\end{document}